\documentclass[aps,prd,nofootinbib,amsmath,amssymb,superscriptaddress,twocolumn,10pt]{revtex4}

\pdfoutput=1

\usepackage{graphicx}
\usepackage{dcolumn}
\usepackage{bm}
\usepackage{amssymb}
\usepackage{latexsym}
\usepackage{booktabs}
\usepackage{amsmath}
\usepackage{multirow}
\usepackage[colorlinks=true, linkcolor=red, citecolor=blue]{hyperref}

\newcommand{\be}{\begin{equation}}
\newcommand{\ee}{\end{equation}}
\newcommand{\bq}{\begin{eqnarray}}
\newcommand{\eq}{\end{eqnarray}}

\begin{document}

\title{Revisiting the holographic dark energy in a non-flat universe: alternative model and cosmological parameter constraints }

\author{Jing-Fei Zhang}
\affiliation{Department of Physics, College of Sciences,
Northeastern University, Shenyang 110004, China}
\author{Ming-Ming Zhao}
\affiliation{Department of Physics, College of Sciences,
Northeastern University, Shenyang 110004, China}
\author{Jing-Lei Cui}
\affiliation{Department of Physics, College of Sciences,
Northeastern University, Shenyang 110004, China}
\author{Xin Zhang\footnote{Corresponding author}}
\email{zhangxin@mail.neu.edu.cn}
\affiliation{Department of Physics, College of Sciences,
Northeastern University, Shenyang 110004, China}
\affiliation{Center for High Energy Physics, Peking University, Beijing 100080, China}

\begin{abstract}

We propose an alternative model for the holographic dark energy in a non-flat universe. 
This new model differs from the previous one in that the IR length cutoff $L$ is taken to be exactly the event horizon size 
in a non-flat universe, which is more natural and theoretically/conceptually concordant with the model of holographic dark energy in a flat universe. 
We constrain the model using the recent observational data including the type Ia supernova data from SNLS3, the baryon acoustic oscillation data 
from 6dF, SDSS-DR7, BOSS-DR11, and WiggleZ, the cosmic microwave background data from Planck, and the Hubble constant measurement 
from HST. In particular, since some previous studies have shown that the color-luminosity parameter $\beta$ of supernovae is likely to vary 
during the cosmic evolution, we also consider such a case that $\beta$ in SNLS3 is time-varying in our data fitting. 
Compared to the constant $\beta$ case, the time-varying $\beta$ case reduces the value of $\chi^2$ by about 35 and results in that 
$\beta$ deviates from a constant at about 5$\sigma$ level, well consistent with the previous studies. For the parameter $c$ of the holographic dark energy, 
the constant $\beta$ fit gives $c=0.65\pm 0.05$ and the time-varying $\beta$ fit yields $c=0.72\pm 0.06$. 
In addition, an open universe is favored (at about 2$\sigma$) for the model by the current data.

\end{abstract}


\maketitle

\section{Introduction}

Current cosmological observations indicate that the expansion of our universe is accelerating due to a mysterious component, called ``dark energy'' \cite{DEReview}, which is gravitationally repulsive and 
dominating the evolution of current universe. To understand the nature of dark energy and explain the observational data, numerous theoretical/phenomenological models of dark energy 
have been put forward during the past 15 years \cite{de1,de2,de3,de4,de5}. Though the nature of dark energy still remains enigmatic, some dark energy models do explain the observational data quite well. 
Besides the famous cosmological constant model or the $\Lambda$ cold dark matter ($\Lambda$CDM) model, which is theoretically challenged, the holographic dark energy model \cite{li04} has been 
attracting lots of attention because it is theoretically plausible and observationally viable. 

In this paper, we focus on the holographic dark energy model based on the holographic principle and an effective quantum field theory. 
According to the energy bound proposed by Cohen et al.~\cite{Cohen:1998zx}, i.e., the total energy of a system with size $L$ would not exceed the mass of a black hole with the same size, 
the vacuum energy density, which is dynamically evolving in such a setting and viewed as the origin of dark energy, called ``holographic dark energy'', is conjectured to be of the form 
\cite{li04}
\begin{equation}\label{eq:rhohde}
 \rho_{\rm de}=3c^2M_{Pl}^2L^{-2},
\end{equation}
where $c$ is a dimensionless parameter, which plays an important role in determining the properties of the holographic dark energy, $M_{Pl}$ is the reduced Plank mass, and $L$ is the IR cut-off length scale of the effective quantum field theory. Initially, Hsu~\cite{Hsu:2004ri}~pointed out that, if $L$ is chosen to be the Hubble scale of the universe, 
the equation of state of dark energy is not correct for describing the accelerating expansion of the universe. 
Then Li \cite{li04} suggested that the IR cut-off $L$ be chosen to be the size of the future event horizon, $R_{h}$, defined as
\begin{equation}\label{eq:rh}
R_h(t)=ar_{\rm max}(t)=a(t)\int^\infty_t{dt' \over a(t')}. 
\end{equation}
This yields a successful model for holographic dark energy, and many theoretical and phenomenological studies followed \cite{hde1,hde2}.  
But this choice is only for a flat universe. 

The next step is to extend the model to a non-flat universe. Huang and Li \cite{huang} considered such an extension, but they did not choose the exact event horizon $R_h$ as the IR cut-off $L$ 
in this case, but took $L$ as
\begin{equation}\label{eq:lr}
 L=ar_{\rm max}(t),
\end{equation}
where
\begin{equation}\label{eq:rk}
r_{\rm max}(t)={1\over \sqrt{k}}{\rm sinn}\left(\sqrt{k}\int_t^{\infty}{dt'\over a(t')}\right),
\end{equation}
 with $\textrm{sinn}(x)=\sin(x)$, $x$, and $\sinh(x)$ for $k>0$, $k=0$,
and $k<0$, respectively. 
Such a non-flat-universe model of holographic dark energy was adopted by the community and led to a number of following-up investigations \cite{khde}. 

In this paper, we propose an alternative model for the holographic dark energy in a non-flat universe. 
Differing from the previous model \cite{huang}, we persist in choosing
the exact event horizon, $R_h$, in a non-flat universe as the IR cutoff $L$ for the holographic dark energy. 
Shown by Weinberg \cite{weinberg}, the event horizon in a non-flat universe is defined as
\begin{equation}\label{eh}
R_{h}(t)=a(t)\int^{r_{\rm max}}_{0}{dr\over\sqrt{1-kr^2}}=a(t)\int^\infty_t{dt' \over a(t')}.
\end{equation}
We argue that this choice is more natural and more theoretically/conceptually concordant with the flat-universe model. 
We will derive the evolution equations for the holographic dark energy in this model setting and test the model with the 
recent observational data including the type Ia supernova (SN) data from SNLS3, the baryon acoustic oscillation (BAO) data 
from 6dF, SDSS-DR7, BOSS-DR11, and WiggleZ, the cosmic microwave background (CMB) data from Planck, and the Hubble constant measurement 
from HST.

We arrange the paper as follows. In Sec.~\ref{sec:2}, we describe the model we propose and derive the evolution equations for the holographic dark energy in a non-flat universe. 
In Sec.~\ref{sec:3}, we describe the observational data we use in the fits. In particular, besides the usual application of the SNLS3 data, we also consider the possibility that the color-luminosity 
parameter $\beta$ is time-varying, which was indicated as an important possibility in recent studies \cite{beta1,beta2,beta3}. We report the fitting results in Sec.~\ref{sec:4} 
and discuss some related issues in Sec.~\ref{sec:5}. 
Summary is given in the final section.

\section{Alternative model of holographic dark energy with spatial curvature}\label{sec:2}

In a spatially non-flat Friedmann-Robbertson-Walker universe, the Friedmann equation can be written as
\begin{equation}
\label{eq:AllOmega} 3M_{Pl}^2 H^2=\rho_{k}+\rho_{m}+\rho_{de}+\rho_{r},
\end{equation}
where $\rho_{k}=-3M_{Pl}^2 k/a^2$ is the effective energy
density of the curvature component, and $\rho_{m}$, $\rho_{de}$,
and $\rho_{r}$ represent the energy densities
of matter (including dark matter and baryons), dark energy, and
radiation, respectively. Define the fractional energy densities of the
various components,
\begin{equation}\label{eq:Omde}
\Omega_{k}={\rho_{k}\over \rho_{c}}, ~~\Omega_{m}={\rho_{m}\over \rho_{c}},~~\Omega_{de}={\rho_{de}\over \rho_{c}}, ~~ \Omega_{r}={\rho_{r}\over \rho_{c}},
\end{equation}
where $\rho_{c}=3M_{Pl}^{2}H^{2}$ is the critical density of the universe. 
The energy conservation equation for the various components in the universe takes the form
\begin{equation}\label{eq:Ec}
\dot\rho_i+3H(1+w_i)\rho_i=0,
\end{equation}
where $w_{1} = -1/3$ for spatial curvature,
 $w_2 = 0$ for nonrelativistic matter, 
 $w_3 = p_{de}/\rho_{de}$ for dark energy, 
 and 
 $w_4 = 1/3$ for radiation. 
Note that, in this paper, an overdot always denotes the derivative with respect
to the cosmic time $t$. Combining  Eqs.~(\ref{eq:AllOmega}) and (\ref{eq:Ec}), we have \cite{Wang:2012uf}
\begin{equation}\label{eq:Pde}
 p_{de}=-\frac{2}{3}\frac{\dot H}{H^2}\rho_c-\rho_c-{1\over3}\rho_{r}+{1\over3}\rho_k.
\end{equation}
Furthermore, this equation, together with the energy conservation equation~(\ref{eq:Ec}) for dark energy, gives
\begin{equation}
\label{eq:OH2}
2(\Omega_{de}-1){\dot H\over
 H}+\dot\Omega_{de}+H(3\Omega_{de}-3+\Omega_k-\Omega_{r})=0.
\end{equation}

In the holographic dark energy model, the most important step is to choose appropriately the IR length cutoff for the effective quantum field theory. 
Different assumptions for the IR cutoff yield various different variants of holographic dark energy \cite{nade,rde,nhde}. 
In the original model of holographic dark energy in a flat universe, $L$ is taken to be the event horizon size of the universe \cite{li04}. 
However, for a non-flat universe, according to Ref. \cite{huang}, $L$ is not taken as the exact event horizon, but the form of Eq.~(\ref{eq:lr}) [along with Eq.~(\ref{eq:rk})] 
is adopted. In this work, we adopt the point of view that the exact form of the event horizon in a non-flat universe should be taken as the IR cutoff for 
the holographic dark energy. Following Weinberg's famous monograph \cite{weinberg} (in which the event horizon in a non-flat universe is clearly defined), we 
take the IR cutoff as 
\begin{equation}\label{eq:rh}
L=R_{h}(t)=a(t)\int^{r_{\rm max}}_{0}{dr\over\sqrt{1-kr^2}}=a(t)\int^\infty_t{dt' \over a(t')}.
\end{equation}

From the definition of the holographic dark energy density (\ref{eq:rhohde}), we have
\begin{equation}
\label{eq:L0} \Omega_{de}={c^2 \over H^2L^2}.
\end{equation}
Substituting Eq.~(\ref{eq:rh}) into Eq.~(\ref{eq:L0}), we get 
\begin{equation}\label{eq:clt}
\int^\infty_{t}{dt \over a}={c\over Ha\sqrt{\Omega_{de}}}.
\end{equation}
Taking the derivative on both sides of Eq.~(\ref{eq:clt}) with respect to $t$ , we can get
\begin{equation}
\label{eq:OL1.3}
{\dot\Omega_{de}\over2\Omega_{de}}+H+{\dot H\over H}={{\sqrt{\Omega_{de}}H}\over c}.
\end{equation}
Combining Eq.~(\ref{eq:OH2}) and Eq.~(\ref{eq:OL1.3}), the two differential equations describing the evolution of holographic dark energy in a non-flat universe can be obtained,
\begin{equation}
\label{eq:deq1}{1\over E}{dE \over dz} =-{\Omega_{de}\over 1+z}
\left({\Omega_k-\Omega_{r}-3\over2\Omega_{de}}+{1\over2} +{{\sqrt{\Omega_{de}}}\over c} \right),
\end{equation}
\begin{equation}
\label{eq:deq2} {d\Omega_{de}\over dz}= -{2\Omega_{de}(1-\Omega_{de})\over 1+z}
\left({{\sqrt{\Omega_{de}}}\over c}+{1\over2}-{\Omega_k-\Omega_{r}\over 2(1-\Omega_{de})}\right),
\end{equation}
where $E(z)\equiv H(z)/H_0$ is the dimensionless Hubble expansion rate,
$\Omega_k(z)=\Omega_{k0}(1+z)^2/E(z)^2$, and $\Omega_{r}(z)=\Omega_{r0}(1+z)^4/E(z)^2$.
In addition, $\Omega_{r0}=\Omega_{m0} / (1+z_{\rm eq})$ with
$z_{\rm eq}=2.5\times 10^4 \Omega_{m0} h^2 (T_{\rm cmb}/2.7\,{\rm K})^{-4}$. 
Here, $h$ is the reduced Hubble constant defined by $H=100h$ km s$^{-1}$ Mpc$^{-1}$, and we take $T_{\rm cmb}=2.7255\,{\rm K}$.
The initial conditions are $E(0)=1$
and $\Omega_{de}(0)=1-\Omega_{k0}-\Omega_{m0}-\Omega_{r0}$.

Furthermore, from the energy conservation equation and the evolution equation of holographic dark energy with spatial curvature, 
using Eq. (\ref{eq:L0}) and Eq.~(\ref{eq:OL1.3}), we can also derive the equation of state for the holographic dark energy in a non-flat universe:
\begin{equation}\label{eq:wde}
w=-{1\over 3}-{2\over 3c}\sqrt{\Omega_{de}}.
\end{equation}
Apparently, this expression of $w$ is the same as the form in a flat universe model \cite{li04}. 
However, one should notice that $\Omega_{de}(z)$ here is determined by the differential equations (\ref{eq:deq1}) and (\ref{eq:deq2}), 
from which the spatial curvature enters.

\section{The observation data}\label{sec:3}
 In this section, we briefly describe the observational data we use in the fits. 


{\it The SN data.}---We use the SNLS3 data compilation \cite{snls3} consisting of 472 data for the combined set of SALT2 and SiFTO. 
Besides the usual application of the SNLS data, here we highlight the consideration of the possibility that the color-luminosity parameter $\beta$ is time-varying 
during the cosmic evolution. It has been shown by the recent studies \cite{beta1,beta2,beta3} that the stretch parameter $\alpha$ is consistent with a constant but the color parameter $\beta$ 
may exhibit significant evolution at high statistical significance (about 6$\sigma$). Thus, besides the consideration of the case of a constant $\beta$, we also consider 
the case of time-varying $\beta$ by using the linear parametrization, $\beta(z)=\beta_{0}+\beta_{1}z$ (note that it has been proven \cite{beta2,beta3} that the evolution of $\beta$ is almost 
independent of the background cosmological model and insensitive to the parametrized form of $\beta$). For the time-varying $\beta$ case, one needs to 
change a small part in procedure, i.e., the total covariance matrix $C=D_{stat}+C_{stat}+C_{sys}$, where $D_{stat}$ is the diagonal part of the statistical uncertainty, 
$C_{stat}$ and $C_{sys}$ are statistical and systematic covariance matrices, respectively. For a more detailed explanation, see, e.g., Refs.~\cite{beta2,beta3}.


{\it The BAO data.}---We use the BAO measurements from several galaxy surveys: 
$r_{\rm s}(z_d)/D_{\rm V}(0.1)=0.336\pm0.015$ from the 6dF Galaxy Survey \cite{Beutler}; $D_{\rm V}(0.35)/r_{\rm s}(z_d)=8.88\pm0.17$ from the SDSS-DR7 \cite{Anderson:2013zyy}; $D_{\rm V}(0.32)\left({r^{\rm fid}_{s}}/{r_s}\right)=1264\pm25~ {\rm Mpc}$ and $D_{\rm V}(0.57)\left({r^{\rm fid}_{s}}/{r_s}\right)=2056\pm20 ~{\rm Mpc}$ from the BOSS-DR11 \cite{Padamanabhan}; $D_{\rm V}(0.44)\left({r^{\rm fid}_{s}}/{r_{s}}\right)=1716\pm83~ {\rm Mpc}$, $D_{\rm V}(0.60)\left({r^{\rm fid}_{s}}/{r_{s}}\right)=2221\pm101~ {\rm Mpc}$, and $D_{\rm V}(0.73)\left({r^{\rm fid}_{s}}/{r_{s}}\right)=2516\pm86~ {\rm Mpc}$ from the ``improved'' WiggleZ Dark Energy Survey \cite{Kazin:2014qga}. 
Note that the three WiggleZ data are correlated with each other, and the inverse covariance matrix for them can be found in Ref.~\cite{Kazin:2014qga}.


{\it The CMB data.}---For the CMB data, we use the ``Planck distance priors'' derived from the Planck first released data~\cite{wangyun}. 
It was shown that the ``acoustic scale'' $l_a\equiv{\pi r(z_*)/ r_{\rm s}(z_*)}$, the shift parameter $R\equiv{\sqrt{\Omega_m H_0^2} \,r(z_*)}$, 
together with the baryon density $\omega_b\equiv \Omega_b h^2$, provide an efficient summary of the CMB data. 
Using the Planck+lensing+WP data and assuming a non-flat universe, the three parameters are obtained: 
$l_a=301.57\pm0.18$, $R=1.7407\pm0.0094$, and $\omega_b=0.02228\pm0.00030$. 
The inverse covariance matrix for them is also given in Ref.~\cite{wangyun}.


{\it The $H_0$ measurement.}---We use the result of direct measurement of the Hubble constant \cite{h0}, $H_{0}=73.8\pm2.0$ km s$^{-1}$ Mpc$^{-1}$, from the supernova magnitude-redshift relation calibrated by the HST observations of Cepheid variables in the host galaxies of eight SN.

We apply the $\chi^{2}$ statistic to estimate the model parameters. For each data set, we calculate 
$\chi_{\xi}^{2}={(\xi^{\rm obs}-\xi^{\rm th})^{2}/\sigma_{\xi}^{2}}$, where 
$\xi^{\rm obs}$ is the measured value of observable given by observation, $\xi^{th}$ is the corresponding theoretic value given by theory, 
and $\sigma_{\xi}$ is the 1$\sigma$ standard deviation. 
In our joint SN+BAO+CMB+$H_0$ fit, the total $\chi^2$ is given by
\begin{equation}
\chi^2 = \chi^{2}_{\rm SN}+\chi^2_{\rm BAO} + \chi^2_{\rm CMB} + \chi^2_{H_{0}}.
\end{equation}
We obtain the best-fit value and the 1--3$\sigma$ confidence level (CL) ranges for the model parameters by performing a Markov-Chain Monte Carlo (MCMC) \cite{cosmomc} likelihood analysis.

\section{The fitting results}\label{sec:4}

\begin{table*}
\caption{The joint constraint results for the dark energy models in non-flat universe, i.e., the $\Lambda$CDM model, the HDE model proposed in this work, and the 
HDE model proposed by Huang and Li \cite{huang}, by using the 
SN+BAO+CMB+$H_0$ data. 
For each model, two cases are considered for the SN data, i.e., the constant $\beta$ and the linear varying $\beta(z)$ cases.}\label{table1}
\begin{tabular}{ccccccccc}
\hline\hline &\multicolumn{2}{c}{$\Lambda$CDM}&&\multicolumn{2}{c}{HDE (this work)}&&\multicolumn{2}{c}{ HDE (HL04 \cite{huang})}\\
           \cline{2-3}\cline{5-6}\cline{8-9}
Parameter  & Constant $\beta$ & Linear $\beta(z)$ && Constant $\beta$ & Linear $\beta(z)$ && Constant $\beta$ & Linear $\beta(z)$ \\ \hline
$c$               &  ...
                   &  ... &
                   & $0.654^{+0.052}_{-0.051}$
                   & $0.721^{+0.063}_{-0.062}$&
                   & $0.644^{+0.057}_{-0.043}$
                   & $0.703^{+0.079}_{-0.042}$\\

$\Omega_{m0}$      & $0.292^{+0.007}_{-0.006}$
                   & $0.296^{+0.006}_{-0.008}$&
                   & $0.281^{+0.008}_{-0.010}$
                   & $0.291^{+0.008}_{-0.010}$&
                   & $0.279^{+0.011}_{-0.009}$
                   & $0.289^{+0.012}_{-0.008}$\\

$10^{3}\Omega_{k0}$
                   & $7.636^{+5.821}_{-5.284}$
                   & $1.582^{+2.401}_{-3.045}$&
                   & $4.902^{+3.024}_{-2.705}$
                   & $7.315^{+3.148}_{-3.463}$&
                   & $1.582^{+2.401}_{-3.045}$
                   & $7.203^{+3.387}_{-2.979}$\\

$h$                & $0.691^{+0.006}_{-0.006}$
                   & $0.690^{+0.006}_{-0.006}$&
                   & $0.707^{+0.013}_{-0.010}$
                   & $0.694^{+0.013}_{-0.010}$&
                   & $0.710^{+0.010}_{-0.013}$
                   & $0.698^{+0.009}_{-0.013}$\\

$\alpha$           & $1.422^{+0.104}_{-0.106}$
                   & $1.410^{+0.111}_{-0.086}$&
                   &$1.425^{+0.010}_{-0.101}$
                   & $1.401^{+0.115}_{-0.072}$&
                   & $1.418^{+0.106}_{-0.094}$
                   & $1.433^{+0.078}_{-0.104}$\\

$\beta_{0}$        & $3.243^{+0.115}_{-0.099}$
                   & $1.441^{+0.323}_{-0.368}$&
                   &$3.251^{+0.112}_{-0.098}$
                   & $1.464^{+0.333}_{-0.347}$ &
                   & $3.273^{+0.084}_{-0.121}$
                   & $1.463^{+0.362}_{-0.328}$\\

 $\beta_{1}$        & ...
                   &$5.092^{+1.052}_{-0.885}$&
                   & ...
                   & $5.057^{+0.943}_{-0.962}$&
                    & ...
                   & $5.028^{+0.938}_{-0.967}$\\

\hline $\chi^{2}_{\rm min}$  & 430.634  & 393.106 && 428.993 & 393.873 && 429.018&393.876 \\
\hline
\end{tabular}
\end{table*}

\begin{figure}
\includegraphics[scale=0.25, angle=0]{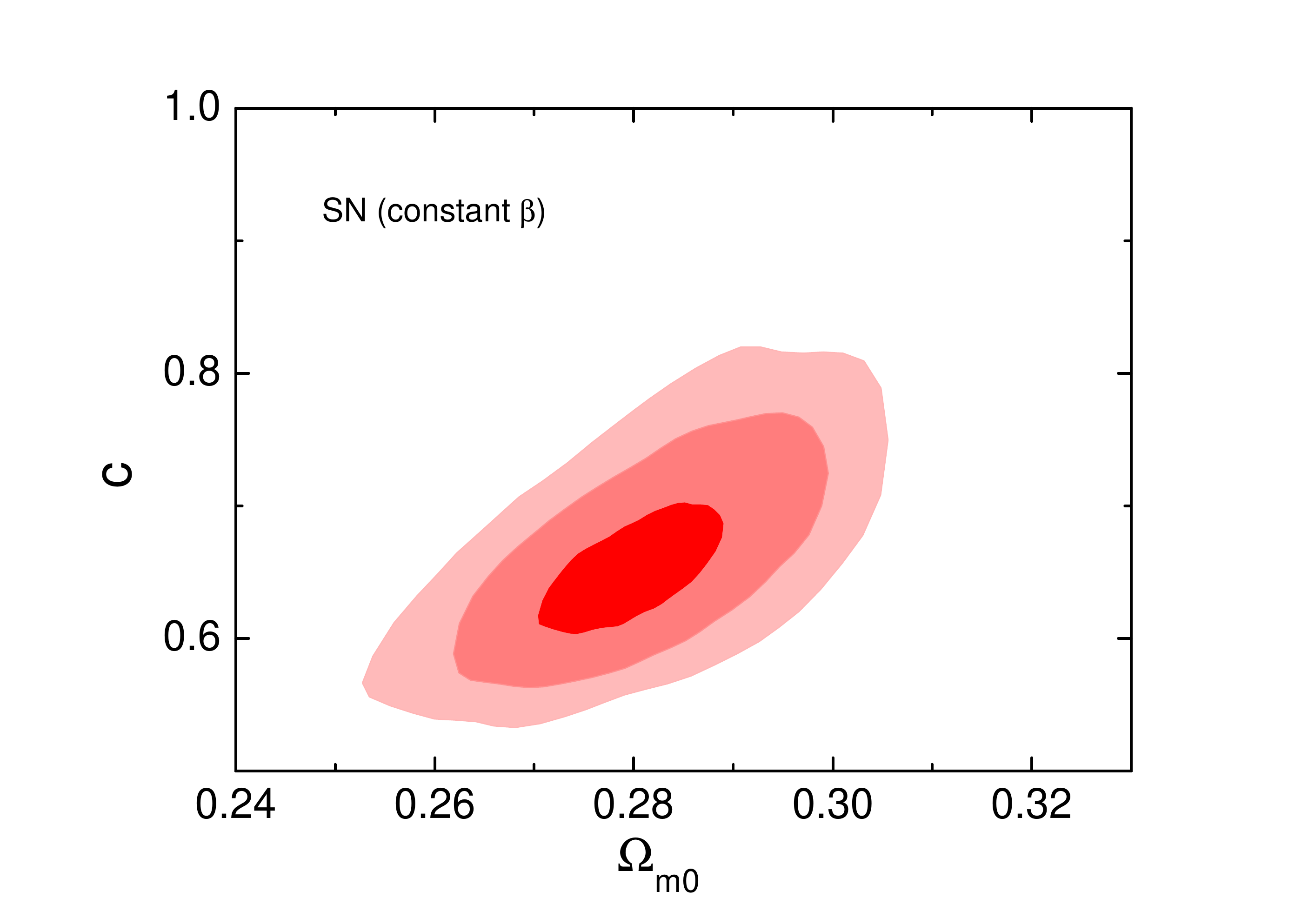}
\includegraphics[scale=0.25, angle=0]{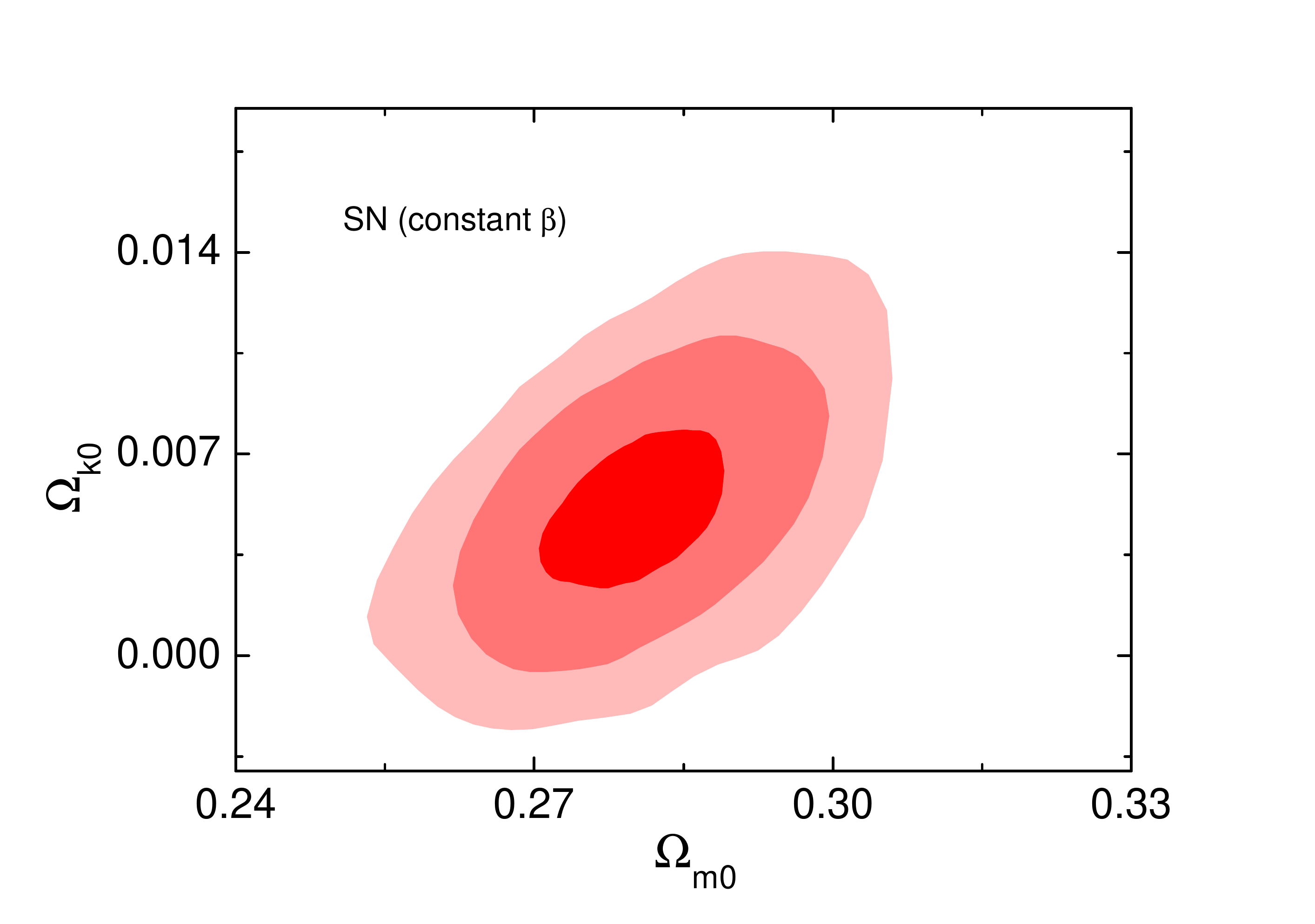}
\caption{\label{fig1}The 1--3$\sigma$ posterior possibility contours in the $\Omega_{m0}$--$c$ and the $\Omega_{m0}$--$\Omega_{k0}$ planes, 
from the SN+BAO+CMB+$H_0$ data, where the constant $\beta$ case for the SN data is considered.}
\end{figure}

\begin{figure}
\includegraphics[scale=0.25, angle=0]{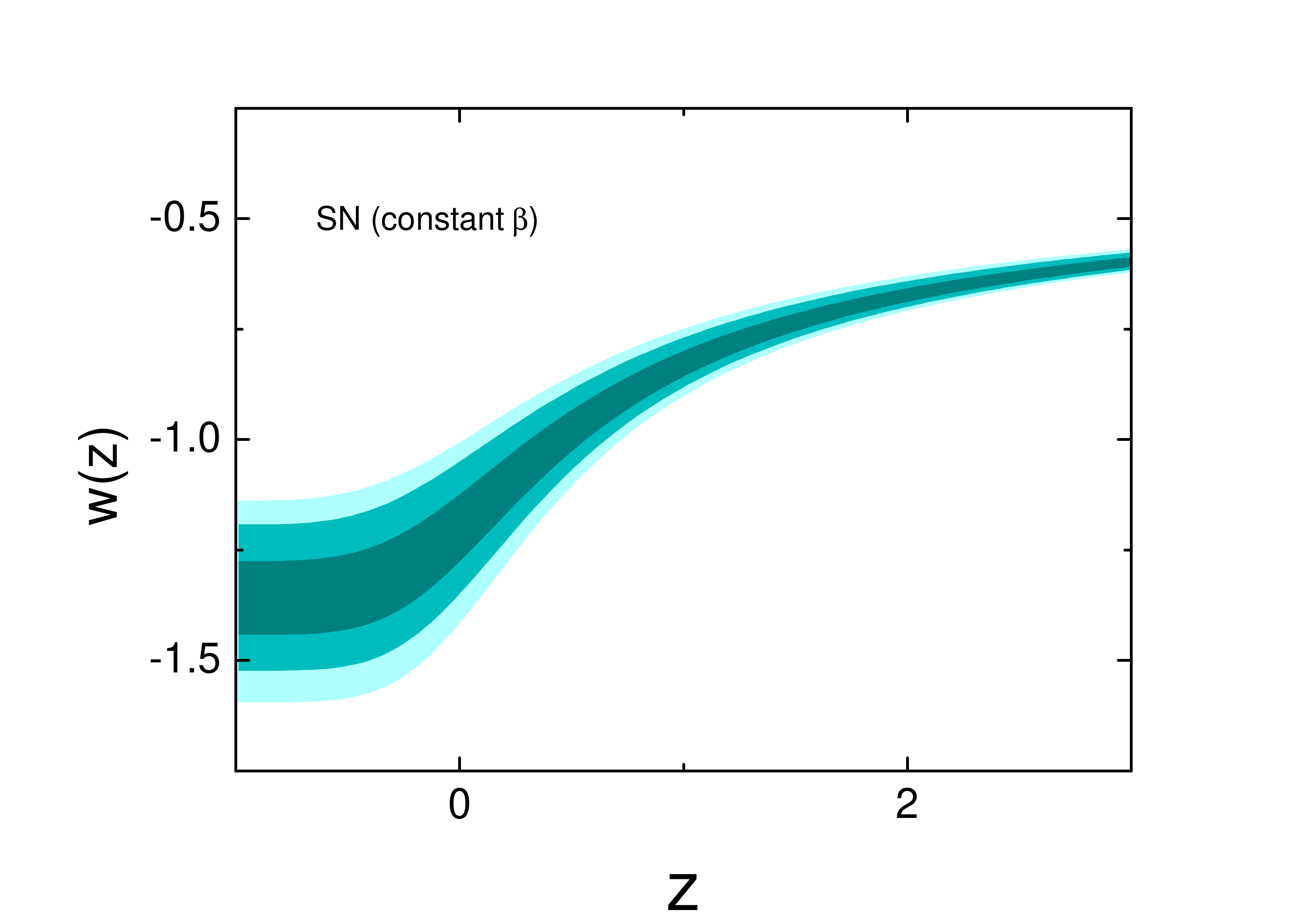}
\caption{\label{fig2}The reconstructed evolution of $w$ under the constraints from the SN+BAO+CMB+$H_0$ data, where the constant $\beta$ case for the SN data is considered.}
\end{figure}

\begin{figure*}
\includegraphics[scale=0.25, angle=0]{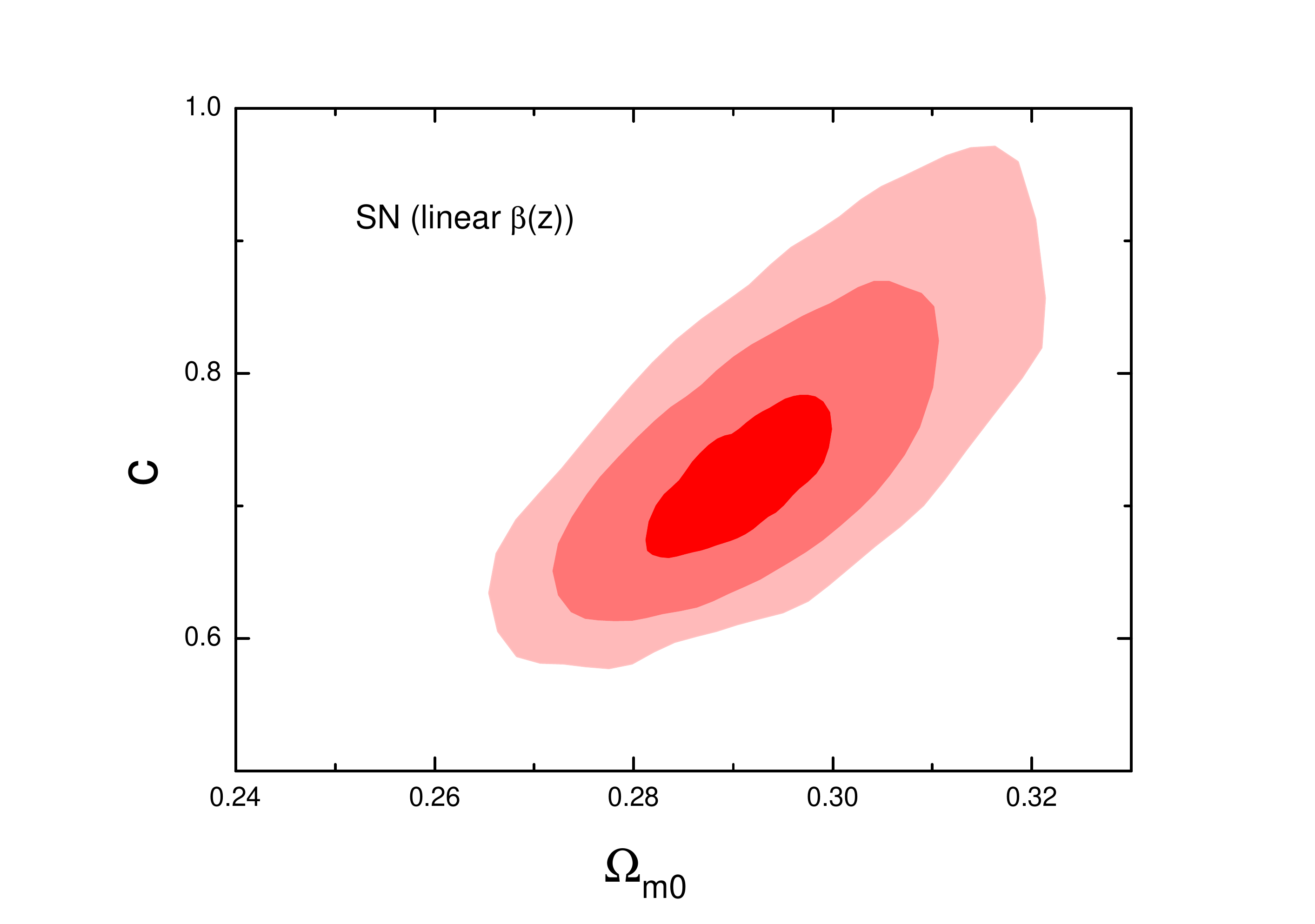}
\includegraphics[scale=0.25, angle=0]{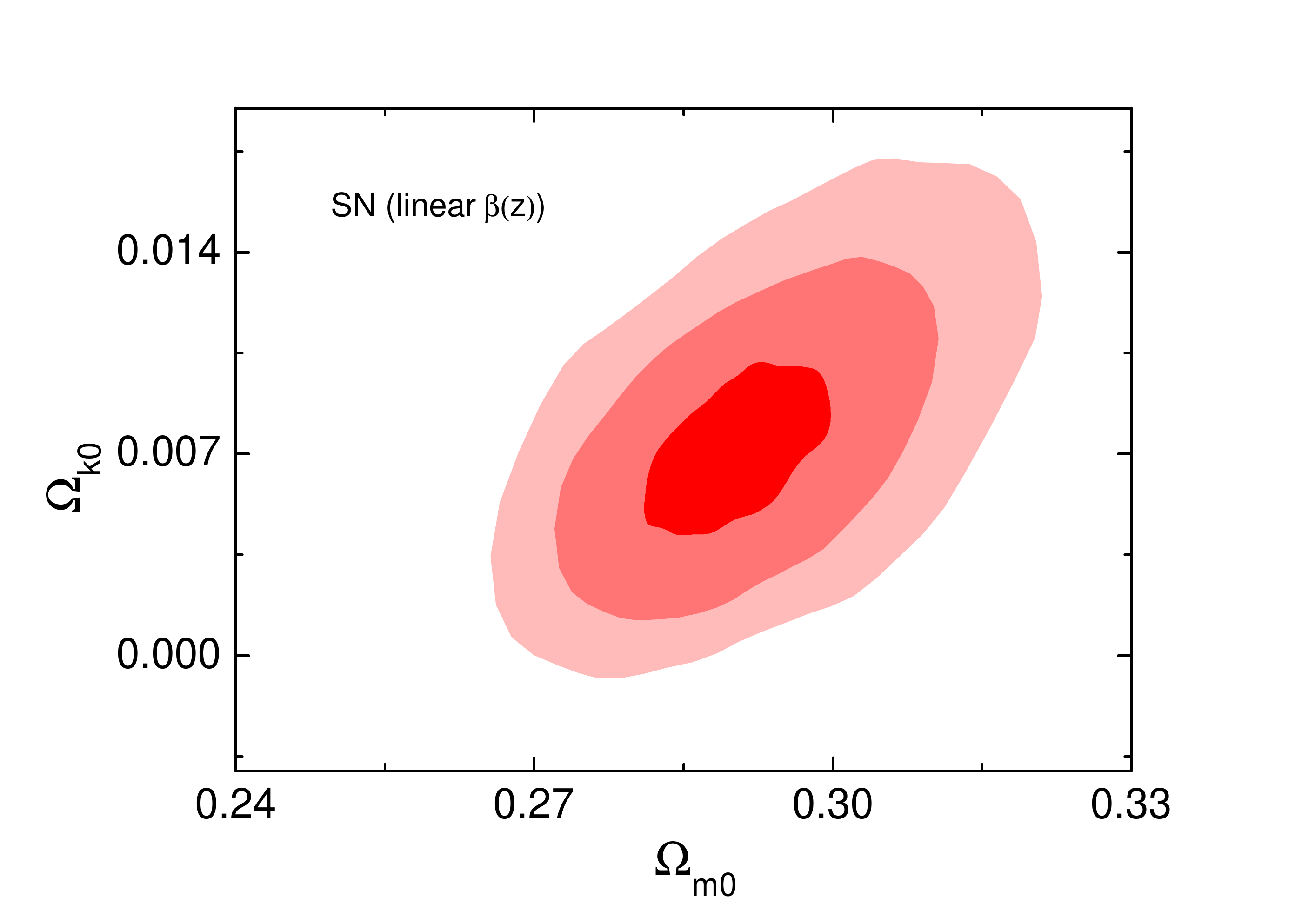}
\includegraphics[scale=0.25, angle=0]{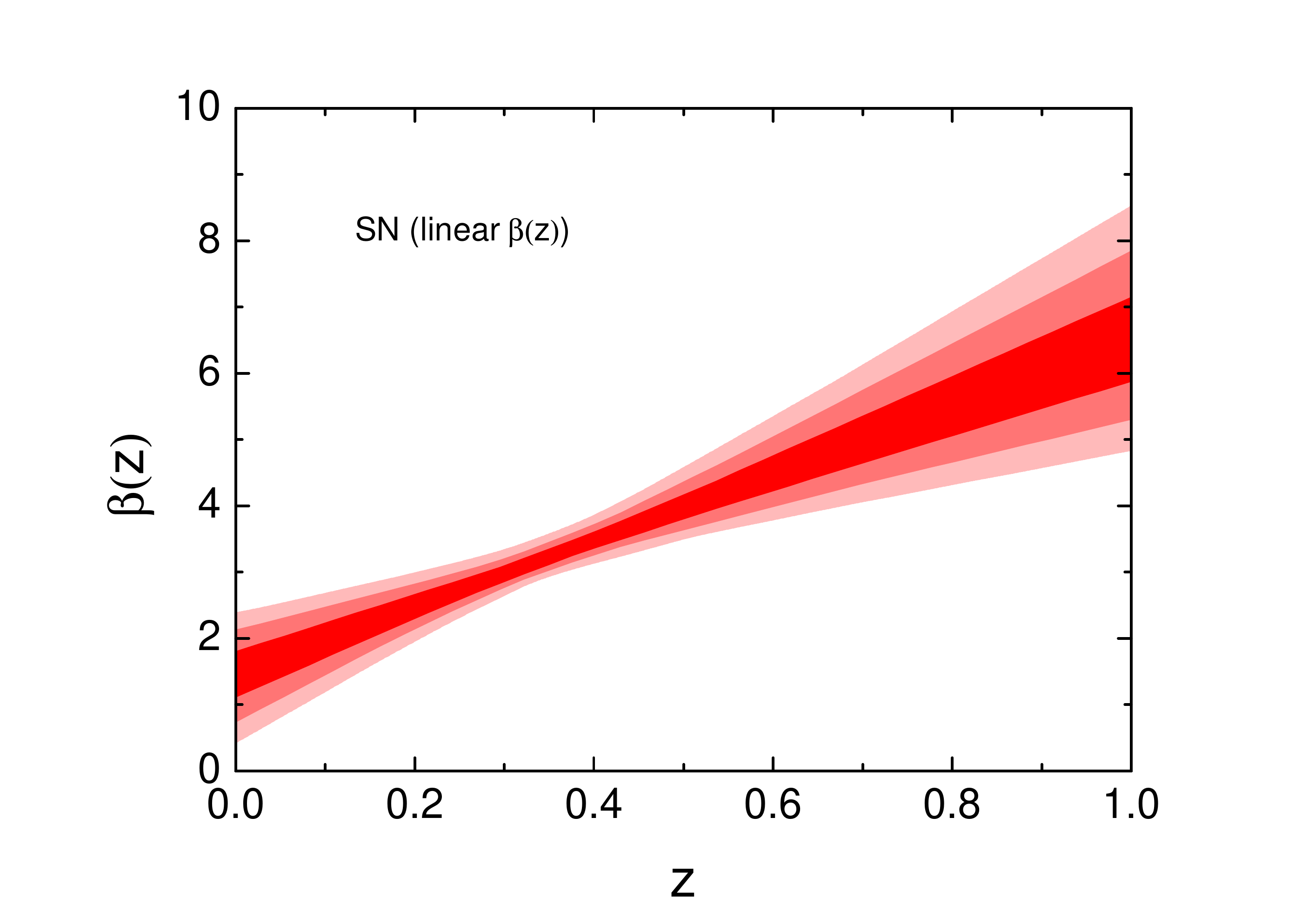}
\includegraphics[scale=0.25, angle=0]{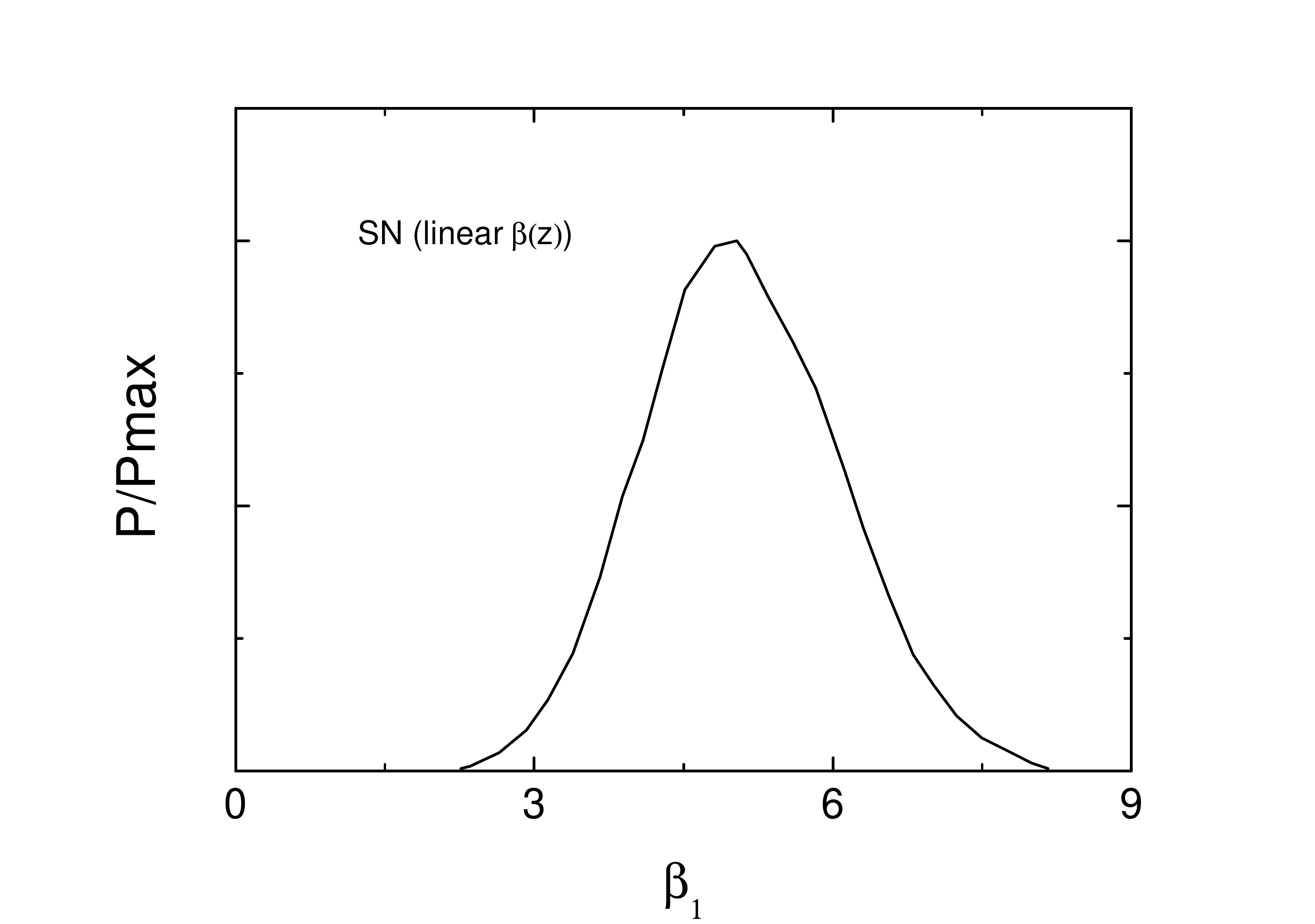}
\caption{\label{fig3}The joint constraints from the SN+BAO+CMB+$H_0$ data, where the linear $\beta(z)$ case for the SN data is considered. 
Upper panels: the 1--3$\sigma$ posterior possibility contours in the $\Omega_{m0}$--$c$ and the $\Omega_{m0}$--$\Omega_{k0}$ planes. 
Lower panels: the reconstructed evolution of $\beta(z)$ and the one-dimensional posterior possibility distribution of $\beta_1$.}
\end{figure*}

\begin{figure}
\includegraphics[scale=0.25, angle=0]{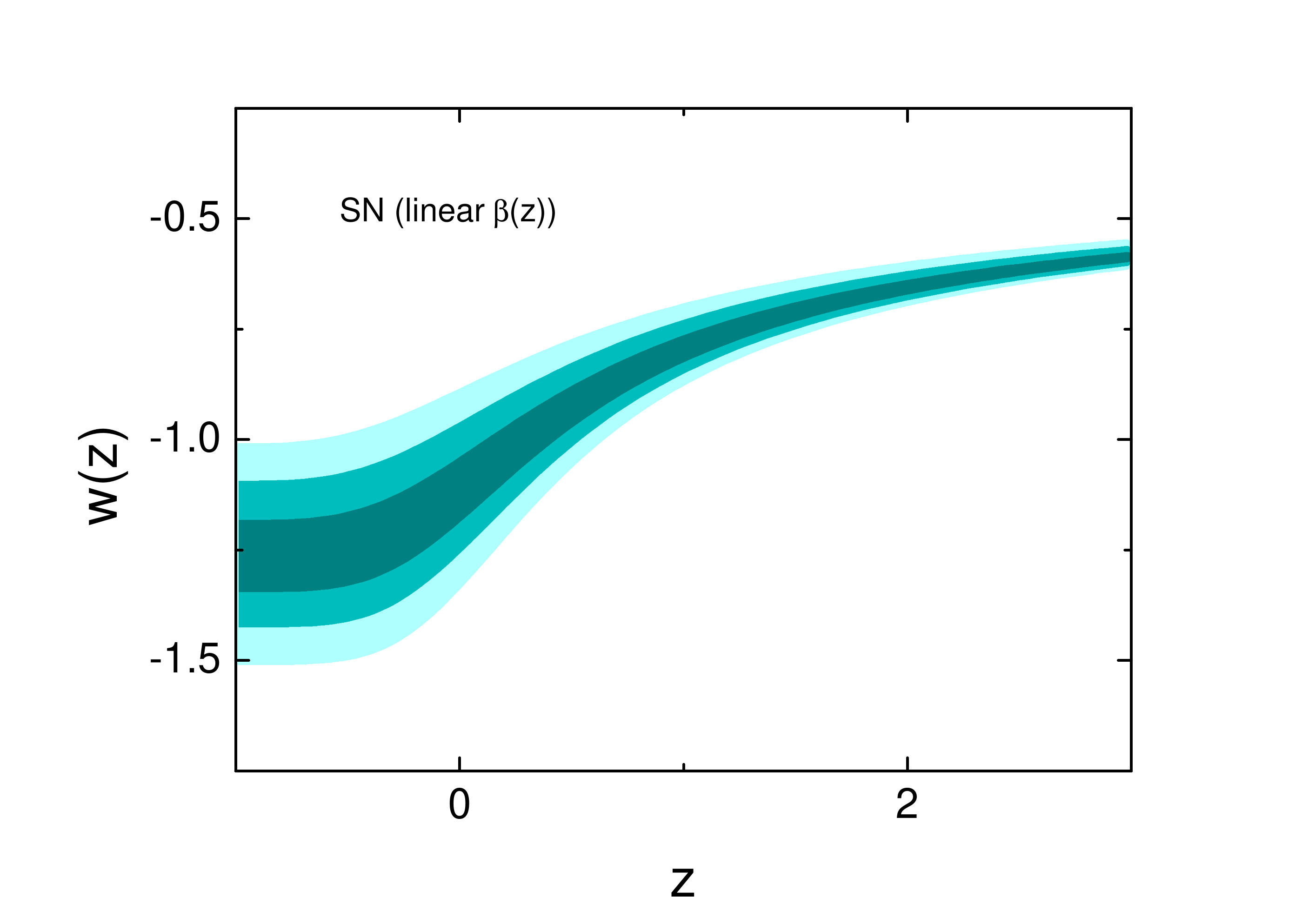}
\caption{\label{fig4}The reconstructed evolution of $w$ under the constraints from the SN+BAO+CMB+$H_0$ data, where the linear $\beta(z)$ case for the SN data is considered.}
\end{figure}

We run eight independent chains with~$300,000 $ data for each chain and obtain the fit values for the cosmological parameters. 
In the joint SN+BAO+CMB+$H_0$ constraints, we consider two cases, i.e., for the SNLS3 data set, we consider constant $\beta$ case 
and time-varying (linear parametrization) $\beta(z)$ case. We will report the fit results for these two cases.

Our constraint results are summarized in Table~\ref{table1}. 
Actually, in this table, we also show the constraint results for the $\Lambda$CDM model and 
the HDE model proposed by Huang and Li \cite{huang} for comparison (here we use HDE as an abbreviation for the holographic dark energy). 
But in this section we will only discuss the results for our model and we 
leave the further discussions including the comparison of the models in the next section.

In Table \ref{table1}, we show the fit values for the important parameters and we 
directly compare the constant $\beta$ case and the linear $\beta(z)$ case. 
The parameters $\alpha$, $\beta_0$, and $\beta_1$ are the parameters of supernova observation. 
Our calculations show that, for the constant $\beta$ case, $\beta=\beta_0=3251^{+0.112}_{-0.098}$, and for the linear $\beta(z)$ case, 
$\beta_0=1.464^{+0.333}_{-0.347}$ and $\beta_1=5.057^{+0.943}_{-0.962}$. 
We find that considering the time-varying $\beta$ can reduce the $\chi^2_{\rm min}$ value by about 35. 
The results are consistent with those obtained in Refs.~\cite{beta2,beta3}. 

We first discuss the results of the constant $\beta$ case. 
The 1--3$\sigma$ posterior possibility contours in the $\Omega_{m0}$--$c$ and the $\Omega_{m0}$--$\Omega_{k0}$ parameter planes 
are shown in Fig.~\ref{fig1}. We obtain the fit results $c=0.654^{+0.052}_{-0.051}$, $\Omega_{m0}=0.281^{+0.008}_{-0.010}$, 
and $\Omega_{k0}=(4.902^{+3.024}_{-2.705})\times 10^{-3}$. 
We find that in this case $c<1$ is at the 6.7$\sigma$ level. Thus, according to this result, the holographic dark energy is very likely to become 
a phantom energy in the future evolution.\footnote{We have derived the equation of state of the holographic dark energy [see Eq. (\ref{eq:wde})], 
$w=-1/3-2\sqrt{\Omega_{de}}/(3c)$. According to this formula, we can easily find that in the early times $w\to -1/3$ (since $\Omega_{de}\to 0$) 
and in the far future $w\to -1/3-2/(3c)$ (since $\Omega_{de}\to 1$). This explains why the phantom divide ($w=-1$) crossing happens when $c<1$.}
To show the $w=-1$-crossing behavior and future phantom manner of the holographic dark energy under 
the current joint constraint, we reconstruct the evolution of equation of state $w$ with 1--3$\sigma$ uncertainties in Fig.~\ref{fig2}. 
In addition, we find that the fit of the holographic dark energy model to the current SN+BAO+CMB+$H_0$ data 
(in the case of constant $\beta$) favors an open universe at the 1.8$\sigma$ level. 

Next, we present the constraint results for the case of linear $\beta(z)$. 
The results of most interest are plotted in Fig.~\ref{fig3}, in which the 1--3$\sigma$ posterior possibility contours in the 
$\Omega_{m0}$--$c$ and the $\Omega_{m0}$--$\Omega_{k0}$ parameter planes are shown in the upper two panels, 
and the lower two panels show the fit result for color parameter $\beta$ of supernova, i.e., the reconstructed evolution of 
$\beta(z)$ (with 1--3$\sigma$ level uncertainties) and the one-dimensional posterior possibility distribution of $\beta_1$. 
We obtain $c=0.721^{+0.063}_{-0.062}$, $\Omega_{m0}=0.291^{+0.008}_{-0.010}$, and $\Omega_{k0}=(7.315^{+3.148}_{-3.463})\times 10^{-3}$. 
So in this case, we find that $c<1$ is at the 4.4$\sigma$ level. Though this still means that the holographic dark energy will become 
a phantom energy in the future, compared to the constant $\beta$ case, the likelihood diminishes evidently. 
We also show the reconstructed evolution of $w$ with 1--3$\sigma$ errors for this case in Fig.~\ref{fig4}, for comparison. 
Moreover, the fit of our model to the current joint data in the case of varying $\beta$ for SN prefers an open universe 
at the 2.1$\sigma$ level. 

Compared to the constant $\beta$ case, the varying $\beta$ case reduces the value of $\chi^2_{\rm min}$ by 35.12, which means that the 
consideration of the evolution of $\beta$ could lead to a much better fit to the data. 
We note that, according to the Akaike information criterion, if $\chi^2_{\rm min}$ improves
by 2 or more with one additional parameter, its incorporation is justified. 
We thus believe that the evolution of $\beta$ perhaps is truly worthy of being considered in the SN treatment. 
We find that in this case $\beta_1$ deviates from 0 at the 5.3$\sigma$ level, as shown in the panel of one-dimensional distribution of $\beta_1$ 
in Fig.~\ref{fig3}. From the present analysis and the previous ones in Refs.~\cite{beta2}, we suspect that the absence of the consideration of $\beta$'s 
evolution perhaps is a potential systematic error source for the supernova data. We have seen that the consideration of time-varying $\beta$ 
in SN data could significantly impact on the joint constraint results.

\section{Discussion}\label{sec:5}

In this section, we discuss some related issues concerning the model presented in this work. 

We first stress the importance of the consideration of the spatial curvature in the holographic dark energy model. 
Actually, the flatness of the observable universe is one of the important predictions of conventional inflationary cosmology. 
The inflation models theoretically produce $\Omega_{k0}$ on the order of the magnitude of quantum fluctuations, i.e., $\Omega_{k0}\sim 10^{-5}$. 
However, the current observational limit on $\Omega_{k0}$ is of order $10^{-3}$ \cite{Ade:2013zuv}. 
On the other hand, since the spatial curvature is degenerate with the parameters of dark energy, it is of great importance to consider the spatial curvature 
in studying dynamical dark energy models \cite{Clarkson:2007bc}. 
Therefore, when we study the holographic dark energy model, in particular, the exploration of the parameter space of the model, it is necessary to include 
$\Omega_{k0}$ as a free parameter in the cosmological fit. 

In this work, we proposed a non-flat universe model for the holographic dark energy. 
Compared to the model by Huang and Li (hereafter, HL04 model) \cite{huang}, the difference is that 
the IR cut-off scale $L$ is taken to be the exact event horizon $R_h$ in our model. 
Our motivation is clear: In the flat-universe model of holographic dark energy, the IR cut-off $L$ is taken to be the event horizon; 
it is obvious that, in the non-flat-universe model, $L$ should also be taken to be the event horizon. 
This is obviously more natural and more theoretically/conceptually concordant with the flat-universe model. 

We also make a comparison with the HL04 model in terms of the results of numerical fit. 
In Table \ref{table1}, we present the fit results for both our model and the HL04 model. 
For the $\chi^2$ values, in the constant $\beta$ case, we obtain $\chi_{\rm min}^2=428.993$ for our model 
and $\chi_{\rm min}^2=429.018$ for the HL04 model, and in the linear $\beta(z)$ case, we obtain 
$\chi_{\rm min}^2=393.873$ for our model and $\chi_{\rm min}^2=393.876$ for the HL04 model. 
We find that our model is only slightly better than the HL04 model in the cosmological fit. 
This is obvious because the difference between the two is rather subtle. 
It should be stressed that the advantage of our model is mainly in the aspect of theoretical consistence. 

Furthermore, the comparison with the $\Lambda$CDM model is also made. 
In Table \ref{table1}, we show the fit results for the $\Lambda$CDM model. 
For the  $\Lambda$CDM model, we have $\chi_{\rm min}^2=430.634$ for the constant $\beta$ case 
and $\chi_{\rm min}^2=393.106$ for the linear $\beta(z)$ case. 
So we find that, in the constant $\beta$ case the holographic dark energy model fits the current data slightly better than 
the $\Lambda$CDM model ($\Delta\chi^2=-1.641$), but in the linear $\beta(z)$ case the holographic dark energy model is slightly 
worse than the $\Lambda$CDM model ($\Delta\chi^2=0.767$). 
Considering that the holographic dark energy model has one more parameter than the $\Lambda$CDM model, the 
latter is actually more favored by the current data. 
This conclusion is in agreement with the previous studies (see, e.g., Ref. \cite{demodels}). 
In fact, the $\Lambda$CDM model is still the best one among various dark energy models in fitting the observational data. 
But we wish to mention that the holographic dark energy model is much better than other related variant models (also with holographic origin), 
e.g., the new agegraphic dark energy model \cite{nade} and the Ricci dark energy model \cite{rde}, in fitting the observational data; 
see Ref. \cite{hdecompare} for an investigation based on the Bayesian evidence and Ref. \cite{demodels} for an investigation 
based on the Akaike and Bayesian  information criteria. 

Finally, we wish to emphasize that the fitting results are insensitive to the parametrized forms of $\beta(z)$.   
In Ref. \cite{beta2}, the authors have tested several parametrization forms for $\beta(z)$, i.e., the linear form, the 
quadratic form, and a step function form, and found that the evolution of $\beta$ and the fitting results are insensitive to 
the forms of $\beta(z)$. 
So in this paper we only consider the linear form of $\beta(z)$ in the cosmological fit.

\section{Summary}

The holographic dark energy in a flat universe is defined by taking the IR cutoff $L$ to be the event horizon of the universe. 
Therefore, to be more theoretically consistent, we put forward in this paper that the holographic dark energy in a non-flat universe should also 
be defined by taking precisely the event horizon of the universe, $R_h$, as the IR cutoff $L$ of the theory. Based on this assumption, we establish an 
alternative model for the holographic dark energy in a non-flat universe, which is, undoubtedly, more conceptually concordant 
with the flat-universe model. 

We then constrain the model by using the recent observational data including the SN Ia data from SNLS3, the BAO data 
from 6dF, SDSS-DR7, BOSS-DR11, and WiggleZ, the CMB data from Planck, and the $H_0$ direct measurement 
from HST. For the SN data, we discuss two cases. Since some previous studies \cite{beta1,beta2,beta3} have shown that the color-luminosity parameter $\beta$ of supernovae is likely to vary 
during the cosmic evolution, besides the constant $\beta$ case, we also consider the case in which $\beta$ is time-varying. 
Owing to the fact that $\beta$ is almost independent of background cosmological model and insensitive to the parametrized form \cite{beta2,beta3}, 
we only consider a linear parametrization form, $\beta(z)=\beta_0+\beta_1z$, in the fits.

We find that, compared to the constant $\beta$ case, the time-varying $\beta$ case reduces the value of $\chi^2_{\rm min}$ by about 35 and results in that 
$\beta$ deviates from a constant at about the 5$\sigma$ level. These results are well consistent with those of previous studies \cite{beta2,beta3}. 
The significant reduction of $\chi^2_{\rm min}$ means that considering the redshift-evolution of $\beta$ could lead to a much better fit to the data. 
We find that the consideration of varying $\beta$ in SN data could largely impact on the results of the joint constraints. 
All these effects we observe might imply that the absence of the consideration of $\beta$'s evolution could be a potential systematic error source 
for the supernova data.

For the parameter $c$ of the holographic dark energy, the constant $\beta$ fit gives $c=0.65\pm 0.05$ (indicating $c<1$ at the 6.7$\sigma$ level) 
and the time-varying $\beta$ fit yields $c=0.72\pm 0.06$ (indicating $c<1$ at the 4.4$\sigma$ level). 
Both cases favor an open universe at about the 2$\sigma$ level.

\begin{acknowledgments}
We thank Jia-Jia Geng, Yun-He Li, and Shuang Wang for helpful discussions.
JFZ is supported by the Provincial Department of Education of
Liaoning under Grant No. L2012087.
XZ is supported by the National Natural Science Foundation of
China under Grant No. 11175042 and the Fundamental Research Funds for the 
Central Universities under Grant No. N120505003.
\end{acknowledgments}

\end{document}